\documentclass[conference]{IEEEtran}

\usepackage{cite}
\usepackage{amsmath,amssymb,amsfonts,stackengine}
\usepackage{algorithmic}
\usepackage{graphicx}
\usepackage{textcomp}
\usepackage{xcolor}
\usepackage{hyperref}
\usepackage{multirow}
\usepackage{longtable}
\usepackage{float}
\usepackage{placeins}
\usepackage{tabularx}
\usepackage{csquotes}
\def\BibTeX{{\rm B\kern-.05em{\sc i\kern-.025em b}\kern-.08em
    T\kern-.1667em\lower.7ex\hbox{E}\kern-.125emX}}

\usepackage{longtable}
\usepackage{booktabs}
\usepackage{soul}

\newcommand{\N}{\mathcal{N}}
\newcommand{\M}{\mathcal{M}}
\newcommand{\bN}{N}
\newcommand{\bM}{M}
\newcommand{\A}{\mathcal{A}}
\newcommand{\B}{\mathcal{B}}

\begin{document}
\title{Predicting the State of Synchronization of Financial Time Series using Cross Recurrence Plots \\
\thanks{Identify applicable funding agency here. If none, delete this.}
}

\author{\IEEEauthorblockN{Mostafa Shabani}
\and
\IEEEauthorblockN{Martin Magris}
\and
\IEEEauthorblockN{George Tzagkarakis}
\and
\IEEEauthorblockN{Juho Kanniainen}
\and
\IEEEauthorblockN{Alexandros Iosifidis}
}

\author{
    \IEEEauthorblockN{
    Mostafa Shabani\IEEEauthorrefmark{1},
    Martin Magris\IEEEauthorrefmark{1}, 
    George Tzagkarakis\IEEEauthorrefmark{2}\IEEEauthorrefmark{3},
    Juho Kanniainen\IEEEauthorrefmark{4},
    Alexandros Iosifidis\IEEEauthorrefmark{1}}
    
    \IEEEauthorblockA{
    \IEEEauthorrefmark{1}Department of Electrical and Computer Engineering, Aarhus University, Denmark
    }
    
    \IEEEauthorblockA{\IEEEauthorrefmark{2}Foundation for Research and Technology - Hellas Institute of Computer Science, Greece }
    
     \IEEEauthorblockA{\IEEEauthorrefmark{3}IRGO (EA 4190), University of Bordeaux, France}

     \IEEEauthorblockA{\IEEEauthorrefmark{4}Department of Computing Sciences, Tampere University, Finland}    
}

\maketitle

\begin{abstract}
Cross-correlation analysis is a powerful tool for understanding the mutual dynamics of time series.
This study introduces a new method for predicting the future state of synchronization of the dynamics of two financial time series. To this end, we use the cross-recurrence plot analysis as a nonlinear method for quantifying the multidimensional coupling in the time domain of two time series and for determining their state of synchronization. We adopt a deep learning framework for methodologically addressing the prediction of the synchronization state based on features extracted from dynamically sub-sampled cross-recurrence plots. We provide extensive experiments on several stocks, major constituents of the S\&P100 index, to empirically validate our approach. We find that the task of predicting the state of synchronization of two time series is in general rather difficult, but for certain pairs of stocks attainable with very satisfactory performance\footnote{Paper submitted to and under consideration at \textit{Pattern and Recognition Letters}.}.
\end{abstract}


\section{Introduction}
Time series prediction and classification in finance is significantly challenging due to the complexity, multivariate nature, and non-stationary nature of time series in this domain\cite{murphy1999technical}. 
Security trading and price dynamics in financial markets are particularly complex due to the interacting nature and inter-connectedness of their underlying driving forces and determinants leading to significant co-movements in stocks' prices. 
The characterization and modeling of multivariate time series dynamics have long been discussed in the financial literature, where 
the prevailing approach is that based on classical econometric theory. 
Among the multivariate linear models, the most widespread ones are vector autoregressive (VAR) models \cite{lutkepohl1999vector,lutkepohl1991introduction}, vector moving averages and ARMA (autoregressive moving average) models \cite{Reinsel1993} and cointegrated VAR models \cite{juselius2006cointegrated}. Widespread is the use of multivariate conditional heteroskedasticity GARCH-type, see e.g. \cite{bauwens2006multivariate} for a review, multivariate stochastic volatility models \cite{harvey1994multivariate}, and more methods based on the realized volatility \cite[e.g.]{chiriac2011modelling}. 

Among the non-linear models the threshold autoregressive model \cite{tong1978tar}, smooth transition autoregressive \cite{dijk2002smooth} and Markov switching models \cite{krolzig2013markov} are nowadays standard approaches. Alternatives include non-parametric methods, functional coefficient \cite{chen1993functional} and nonlinear additive AR models \cite{chen1993nonlinear}, recurrence analysis, and neural networks.
The complexity of modern financial markets running over the so-called limit-order book mechanism is, however, characterized by typical non-linear, noisy, and often non-stationary dynamics. In addition, the high-dimensional nature of the limit-order book flow and complexity of the interactions within it constitute severe limits in the applicability of classic econometric methods for its modeling and forecasting. Besides a very limited number of analytical and tractable models for the order flow and price dynamics in limit-order books \cite[e.g.]{cont2010stochastic,huang2012generalized,hawkes2018hawkes}, machine learning methods have received much attention  \cite{heaton2017deep,dixon2020machine}, as they are naturally appealing in this context. 

By considering the stock market as a complex system, it is natural to apply such methods for addressing those prediction problems where the application setting and assumptions beneath standard econometric techniques are stringent or inadequate. 
Indeed, it has extensively been shown that, in financial applications, deep learning (DL) models are often capable of outperforming traditional approaches due to their ability to learn complex data representations based on end-to-end data-driven training, see e.g. \cite{Sezer2017, Zhang2019, tran2018temporal, passalis2020temporalBoFs, shabani2022augmented,HASELBECK2022100239}. DL models have been adopted for a variety of problems ranging from price prediction 
\cite{khare2017short,fons2021augmenting,BHANDARI2022100320,BASHER2022100355,XU2021100140}, limit-order book-based mid-price prediction \cite{Zhang2019, tran2018temporal, shabani2022augmented, shabani2020low, shabani2022multi}, and volatility prediction \cite{kyoung2019performance, liu2019novel,christensen2021machine}.

Whereas the target of the above literature is generally the analysis and prediction of single time series, this paper focuses on an analysis of stock pair co-movements. Several trading strategies can be put into play to take advantage of co-movements and exploit statistical arbitrages, including pair-trading, portfolio management, or relative and convergence trading strategies applied at an intraday level, e.g., see \cite{guo2017quantitative} for an overview. While DL provides a basis for prediction given a set of descriptive features, the issue of how to detect and quantify co-movements remains to be addressed. This paper suggests the use of recurrence analysis based on Cross Recurrence Plots (CRP) for detecting and extracting features indicative of stocks' shared dynamics or co-movements, along with a deep learning framework for predicting whether certain pairs of stocks will exhibit a shared dynamics in the future (in the sense specified in Section \ref{S:RecurrenceAnalysis}).
Not only in the view of extending the ML and applied econometrics literature in this direction, but the possibility of forecasting epochs of time series synchronization is likewise relevant for practitioners.

For detecting and quantifying co-movements or more generally shared dynamical features in time series, the standard econometric approach is that of cross-correlation analysis, e.g., \cite[Chapter 8]{tsay2005analysis}. This intuitive linear approach, based on the estimation and perhaps forecasting of cross-correlation matrices, appears to be an element of a much wider theory and methodological approach that has been explored and developed in the last years within a broader generic non-financial setting. Simple cross-correlation analysis has been remarkably extended and generalized towards methods that help explore co-movements between time series within non-linear, noisy, and non-stationary systems of very complex dynamics, either financial \cite{ma2013multifractal, bonanno2001high, ramchand1998volatility} or not \cite{webber1994dynamical,marwan2009nonlinear,lancia2014application}.

Recurrent analysis \cite{webber2015recurrence} explores the reconstruction of a phase-space using time-delay embedding for quantifying characteristics of nonlinear patterns in a time series over time \cite{takens1981detecting}. This is done by calculating the so-called Recurrence Plot  \cite{eckmann1995recurrence}, the core concept of which is to identify all points in time that the phase-space trajectory of a single time series visits roughly the same area in the phase-space. Recurrence plot analysis has no assumptions or limitations on dimensionality, distribution, stationarity, and size of data \cite{webber2015recurrence}. These characteristics make it suitable for multidimensional and non-stationary financial time series data analysis. 
The CRP \cite{marwan1999examination} is an extension of the recurrence plot, introduced to analyze the co-movement and synchronization of two different time series. The CRP indicates points in time that a time series visits the state of another time series, with possibly different lengths in the same phase-space. These concepts are discussed in further detail in Section \ref{S:RecurrenceAnalysis}.

In this paper, we propose a method for predicting the state of synchronization over time of two multidimensional
\footnote{Throughout the paper, with uni- or multi- \textit{variate} we refer to the nature of the analyses (RP as opposed to CRP), and with one- or multi- \textit{dimensional} we refer to the nature of the time series. That is, the RP (as presented in equation \eqref{eq:R}) provides a univariate analysis of a single one-dimensional time series, while the CRP (as presented in latter equation \eqref{eq:CRij}) a multivariate analysis of two one-dimensional or multi-dimensional time series.} 
financial time series based on their CRP. 
In particular, we use the CRP to quantify the co-movements and extract the binary representation of its diagonal elements as the prediction targets for a DL model.
For predicting the state of synchronization at the next epoch we employ a Convolutional Neural Network (CNN) that uses as inputs CRPs independently calculated from data-crops obtained by applying fixed-size sliding windows on the time series.
Our extensive experiments on 12 stocks of the S\&P100 index selected from different sectors show that the proposed method can predict the synchronization of stock pairs with satisfactory performance. 

The remainder of the paper is organized as follows.
Section \ref{S:RecurrenceAnalysis} introduces in detail the concepts and theory behind the CRP, with an outlook on its applications in financial and economic problems. Our proposed approach for predicting time instances of time series' synchronization is presented in Section \ref{S:Method}. Empirical results on real market data are provided in Section \ref{S:Experiments}, whilst Section \ref{S:Conclusion} provides conclusions.

\section{Financial Time Series Recurrence Analysis}\label{S:RecurrenceAnalysis}
Recurrence in the analysis of time series, seen as a nonlinear dynamic system, is the repetition of a pattern over time. The visualization of recurrences in the dynamics of a time series can be expressed via a RP or recurrence matrix\cite{webber2015recurrence}. In other words, the RP represents the recurrence of the phase-space trajectory to a state. The phase-space of a $d$-dimensional time series $\mathcal{N}$ with $T$ observations $\mathcal{N} =\{\mathbf{n}_1^\top, \mathbf{n}_2^\top,\dots, \mathbf{n}_T^\top\}^\top$, with $\mathbf{n}_i$ being the row-vector representing a generic observation at time $i$, $i = 1,\dots,T$ is calculated using the time-delay embedding method. State $\bN_i$ in the phase-space is obtained by
\begin{equation}\label{eq:X_i}
    N_i = [\mathbf{n}_i, \mathbf{n}_{i+\tau},\dots, \mathbf{n}_{i+(k-1)\tau}], \:\: i=1,\dots,T' \text{,}
\end{equation}
where $\tau$ denotes the delay and $k$ is the embedding dimension, $T' =  T-\tau (k-1)$,
$\tau$ and $k$ can, respectively, be determined with the Average Mutual Information Function (AMI) method\cite{fraser1986independent} and the False Nearest Neighbors (FNN) method of \cite{kennel1992determining}.
For a uni-dimensional time series $N_i$ is a row vector of size $(1 \times k)$, for a $d$-dimensional times series $N_i$ is a row vector of size $(1 \times kd)$.
The recurrence state matrix of the reconstructed phase-space, known as Recurrence Plot (RP), at times $i$ and $j$, is defined as
\begin{equation}\label{eq:R}
    R_{i,j}(\varepsilon) = H(\varepsilon - \|\bN_i - \bN_j\|), \:\: i,j = 1,\dots,T' \text{,}
\end{equation}
where $\varepsilon$ is a threshold distance value, $H(\cdot)$ is the Heaviside function, and $\| \cdot \|$ is the euclidean distance.
Due to the underlying embedding \eqref{eq:X_i}, $R_{i,j}$ is defined for i $i = 1$ up to $T' = T-\tau (k-1)$. 
If two states $\bN_i$ and $\bN_j$ are in an $\varepsilon$-neighbourhood the value of $R_{i,j}$ is equal to $1$, otherwise is $0$.The value of $\varepsilon$ highly affects the output of RP. When $\varepsilon$ is too small or too large, the RP cannot identify the true recurrence of states. There are different approaches for finding the best value for $\varepsilon$ in the literature \cite{webber2015recurrence}. We follow the guidelines provided in \cite{schinkel2008selection} for selecting $\varepsilon$. The values on the diagonal line of RP are equal to one (i.e., $R_{i,i} = 1$) because in that case the two states introduced to $H(\cdot)$ are identical. The diagonal line of RP is called the Line Of Identity (LOI).
Recurrence quantification measures derived from RPs, such as recurrence rate (RR), percent determinism (DET), and maximum line length in the diagonal direction (Dmax) \cite{webber2015recurrence}, give  insights about the dynamic behavior of time series. These measures have been used in financial research to analyze the behavior of financial data, e.g. \cite{bastos2011recurrence, fabretti2005recurrence, yin2016multiscale, holyst2001observations, zbilut2005use}. The RP of a time series can be used as a data transformation method for time series prediction. A method employing the RP of seven financial time series to train a deep neural network for predicting the market movement is proposed by \cite{hailesilassie2019financial}. 
Several authors have suggested an RP forecasting approach via DL. A feature extraction method exploiting the RP for parsing a DL algorithm is proposed by \cite{li2020forecasting}. On the other hand, RP can be treated as images enabling the use of different computer-vision techniques for the forecasting task, e.g. \cite{sood2021visual7} uses autoencoders or \cite[e.g.]{han2021identification} uses a CNN. 

The CRP of two multi-dimensional time series \cite{wallot2019multidimensional} corresponds to an extension of RP which explores the co-movement of two time series, and allows the study of the non-linear dependencies between them. Let us denote by $\bN_i, \:i=1,\dots,T$ and $\bM_j, \:j=1,\dots,S$ the phase-space states of the time series $\N$ and $\M$ of length $T$ and $S$, respectively. The Cross-Recurrence (CR) states of the reconstructed state-space a time $i$ and $j$ are calculated by
\begin{equation}\label{eq:CRij}
    CR_{i,j}(\varepsilon) = H(\varepsilon - \| \bN_i - \bM_j \|) 
\end{equation}
with $i = 1,\dots,T' = T-\tau(k-1)$ and $j=1,\dots,S' = S-\tau(k-1)$. Here $\bN_i$ and $\bM_i$ are defined as in \eqref{eq:X_i}.
$CR_{i,j}$ defines the concept of synchronization and the way synchronization between two financial time series measured: an $\varepsilon$-neighbourhood of the embeddings $\bN_i$, $\bM_j$ at epochs $i,j$.
We denote the full cross-recurrence matrix, known as cross-recurrence plot (CRP) extracted for $\N$ and $\M$ as $CRP_{(\N,\M)}$, obtained through
\begin{equation}
CRP_{(\N,\M)} := \left\{ CR_{i,j}(\varepsilon) \right\}_{i = 1,\dots,T',\, j = 1,\dots,S'}\text{.} 
\end{equation}
The CRP corresponds to a matrix of dimension $T' \times S'$, which may not be square, as the time series $\mathcal{N}$ and $\mathcal{M}$ may have different lengths, i.e., $T\neq S$. 

For $\N$, $\M$ of equal length $T$ the $CRP_{(\N,\M)}$  is a square $T'\times T'$ matrix. Opposed to the (univariate) recurrence analysis of one time series (with itself, RP in equation \eqref{eq:R}), the diagonal entries of the CRP are either 1 or 0, as the two time series may or may not be synchronized at $(i,j)$, $i=j, i = 1,\dots,T'$, see e.g. Figure \ref{fig:CRP_sample}. 
In the CRP of two time series the LOI is replaced by a distorted diagonal, called the Line Of Synchronization (LOS). The LOS reveals the relationship between the two time series in the time domain. In particular, it provides a non-parametric function containing information about the time-rescaling of the two time series, that further allows their re-synchronization \cite{marwan2002cross}.

As the time series we consider in our application are multidimensional ($d>1$), we point out that the CRP is indeed a Multidimensional Cross-Recurrence Plot (MdCRP) \cite{wallot2019multidimensional} where $\mathbf{n}_i$, $\mathbf{m}_j$ are row-vectors rather than scalars, and $\bN_i$, $\bM_j$ are $dk$-dimensional row vectors rather than $k$-dimensional row vectors, as opposed to the conventional CRP based on one-dimensional time series. Yet the above discussion is general and applies to both cases, and $CR_{i,j}$ is in any case a scalar equal to either 0 or 1. For multidimensional time series, the entries of each of the two time series require normalization in each dimension before estimating the MdCRP \cite{wallot2019multidimensional}, e.g. with the $z$-score. 

Financial time series co-movement analysis using the CRP and LOS is studied in \cite{crowley2008analyzing, guhathakurta2014understanding, he2020method}. 
\cite{tzagkarakis2016restoring} analyses the inter-dependencies of the stock market index and its associated volatility index, further proposing a method for the LOS estimation based on a corrupted CRP. 

Furthermore, it is important to notice that financial time are often represented as multivariate instance. Indeed, the most basic source of financial data generally provide information about volumes along with prices. 
Despite the use of multidimensional inputs being effective and commonly found across several applications \cite{webber2015recurrence}, existing CRP applications on market data are broadly limited to the use of one-dimensional series only (e.g. prices or volatilities) \cite{webber2015recurrence,orlando2018recurrence,addo2013nonlinear,bastos2011recurrence}.

\section{Proposed Method}\label{S:Method}
\begin{figure*}[!htb]
\centering
\includegraphics[width=0.75\textwidth]{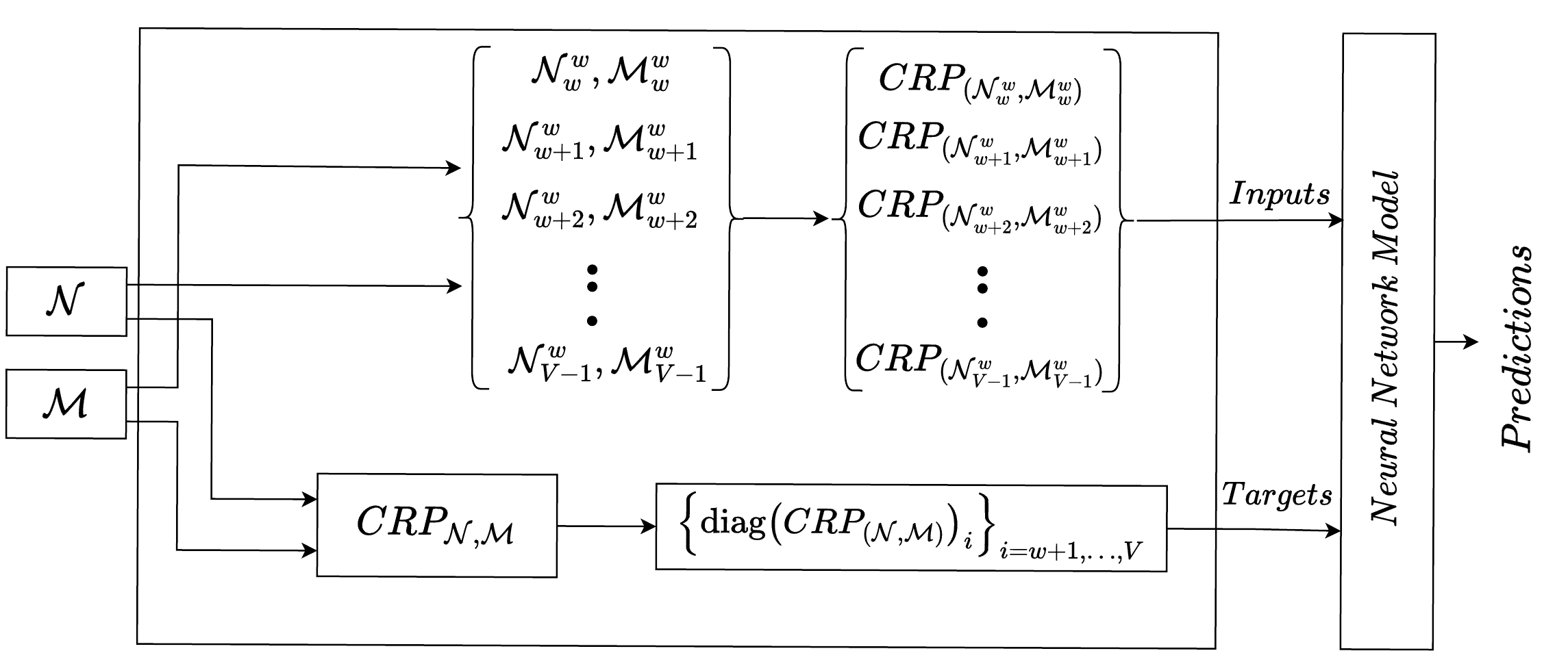}
\caption{Proposed method.}
\label{fig:method}
\end{figure*}

We exploit the CRP to quantify the co-movement of two multidimensional time series $\N$ and $\M$ observed continuously over a common period of length $V$.
Our goal is to predict whether at a certain epoch (e.g. a certain calendar day), $\N$ and $\M$ are synchronized in the state-space embedding $\varepsilon$-neighbourhood given by \eqref{eq:CRij}.

Assume two generic time series $\N$ and $\M$ are observed over the non-overlapping time-domains $D_\N = \{ t_1,\dots,t_T\}$ and $D_\M = \{ s_1,\dots,s_S\}$ of respective length $T$ and $S$, their CRP corresponds to a $T' = (T-\tau (k-1)) \times (S-\tau (k-1)) = S'$ matrix where the $(i,j)$ element expresses the state of synchronization at the $i$-th time instance of the first time series ($t_i$) and at the $j$-th time instance of the second ($s_j$), in terms of the $\varepsilon$-neighbourhood of the states $\bN_i$ and $\bM_j$, as expressed by \eqref{eq:CRij}.
With the domains being non-overlapping there are no epochs $t_i$ and $s_j$ such that $t_i = s_j$ in (the same) calendar time, and the state of synchronization at a same calendar time cannot be determined. Indeed, for a fixed $t_i$, the column-vector $CRP_{i,\cdot}$ reports for the epochs $s_j,\, j=1,\dots,S'$ (past or future with respect to $t_i$) whether state-space embedding $\bM_j$ is in the same $\varepsilon$-neighbourhood of $\bN_i$.

In this light, if the domains of $\N$ and $\M$ only partially overlap over a region $D:=D_{\N} \cap D_{\M}$ of length $V$, for our forecasting purpose their non-overlapping regions $D_\N \setminus D$ and $D_\M \setminus D$ are irrelevant and can be discarded. 
On the other hand, over $D$, their $V$ overlapping time instances $t_i,\dots,t_{i+V}$ and $s_j,\dots, s_{j+V}$ correspond to the same calendar epochs, i.e. $t_{i+h} = s_{j+h},\, \forall h = 1,\dots,V$, and are of actual relevance. 
Over the common domain $D$, the CRP corresponds to a square $V' \times V'$ matrix ($V' = V-\tau (k-1)$) with a well-defined diagonal expressing the state of synchronization at $t_i = s_j$, e.g. answering whether at the (same) calendar day $t_i = s_j$, $\N$ and $\M$ are synchronized or not.

This justifies the required form for the input data, corresponding to two (multidimensional) time series $\N$ and $\M$ observed over a common period $D$ of equal length $V$, with $D =\left\{ v_1 = \text{max}(t_1,s_1),\dots, v_V = \text{min}(t_T,s_S)\right\}$.
Since the essence of time series forecasting is that of predicting the future from the past, the data from the past needs to be representative for the $h$-step ahead forecast. Trivially, this implicitly requires $D$ to be a continuous set of times for the given sampling frequency. That is, there should be no gaps between days or epochs, namely, $v_V \equiv v_1 + (V-1)$.
Furthermore, in order to calculate \eqref{eq:CRij}, we require the time-series to be non-corrupted over $D$ in all its multivariate entries, i.e. without missing values. 

The above requirements are generally met for the financial time series of our interest. The only constraints are that of using data for stocks traded at the same exchange (same trading days and observed festivities), and that of selecting stocks not subject to delisting in the period of interest. As a precaution, we suggest inspecting the data for missing values due to data-quality issues related to the data provider, or unlikely security-related events such as trading halts. 


Aligned with the general rationale of time series forecasting, we aim at predicting the one-step-ahead synchronization status between $\N$ and $\M$ at epoch $i+1$, based on some lagged historical records available up to time $i$, that is based on some suitable set of feature observed or extracted over $w$ past epochs. 
For $i = w,\dots,V'-1$ let us denote by $\N^w_i$ and $\M^w_i$ the sub-sample of $\N$ and $\M$ of the $w$ most recent observations up to and including epoch $i$, that is
\begin{align*}
 \N^w_i &= [ \mathbf{n}_{i-w+1}, \mathbf{n}_{i-w+2},\dots, \mathbf{n}_{i}]\text{,}\\
  \M^w_i &= [ \mathbf{m}_{i-w+1}, \mathbf{m}_{i-w+2},\dots, \mathbf{m}_{i}]\text{.}
\end{align*}
Let us denote by $CRP_{(\N^w_i ,\M^w_i)}$ the $w'\times w'$ CRP computed from $\N^w_i$ and $\M^w_i$ (with embedding dimension $k$, lag $\tau$ and $w' = w-\tau(k-1)$). At epoch $i$, $CRP_{(\N^w_i ,\M^w_i)}$ is used as the input of the neural network for predicting the state of synchronization at $i+1$. Within this framework, there are $V'-w$ input-target pairs. The first pair corresponds to the input $CRP_{(\N^w_{w},\M^w_{w})}$ and target $(CRP_{(\N,\M)})_{w+1,w+1}$, the last to the input $CRP_{(\N^w_{V'-1},\M^w_{V'-1})}$ and target $(CRP_{(\N,\M)})_{V',V'}$.
The prediction target at epoch $i$ corresponds to the state of synchronization at $i+1$, provided by the (diagonal) entry $\left(CRP_{(\N,\M)}\right)_{i+1,i+1}$ of the CRP computed for the entire times series $\N$, $\M$.
In particular, the state of synchronization at any epoch $i=1,\dots,V'$ is provided by the diagonal of $CRP_{(\N,\M)}$, i.e.,
\begin{equation}
\underset{i= w,...,V'-1}{\mathbf{diag}(CRP_{(\N, \M)})_i}=
    \begin{cases}
      1 & \stackunder{\text{if} $\N$ and $\M$}{\text{are synchronized at time $i$},}\\
      0 & \text{\;\;\;\;\;\;\;\;\;\;\;otherwise,}
    \end{cases}
\end{equation}
so that $\left\lbrace\text{diag}\left(CRP_{(\N,\M)}\right)_i \right\rbrace_{i=w+1,\dots,V'}$ corresponds to the targets for the inputs $\left\lbrace  CRP_{(\N^w_{i} ,\M^w_{i})} \right\rbrace_{i = w,\dots,V'-1}
$. 
The above corresponds to a framework where inputs are created dynamically by using CRPs computed over sub-sampled time series obtained by applying sliding windows of a fixed size.
Note that $CRP_{(\N^w_i,\M^w_i)}$ is not analogous to the sub-matrix $P$ obtained from $CRP_{(\N,\M)}$ by considering its rows and columns from $i-w+1$ to $i$. In $CRP_{(\N,\M)}$ the entire data in $\N$ and $\M$ accounts for the time series normalization and furthermore tunes the parameter $\varepsilon$. $CRP_{(\N^w_i,\M^w_i)}$ is thus truly dependent on the cropped times series data $\N^w_i$, $\M^w_i$, while $P$ is not. In a forecasting context our approach is feasible and unbiased as it does not use any future information following the one available at $i$.
Note that, in general, nothing prevents from choosing the embedding size and lag parameter differently for the CRP computation of the targets and for the CRP computations of the inputs.

\begin{figure}[!htb]
\centering
\includegraphics[width=0.75\columnwidth]{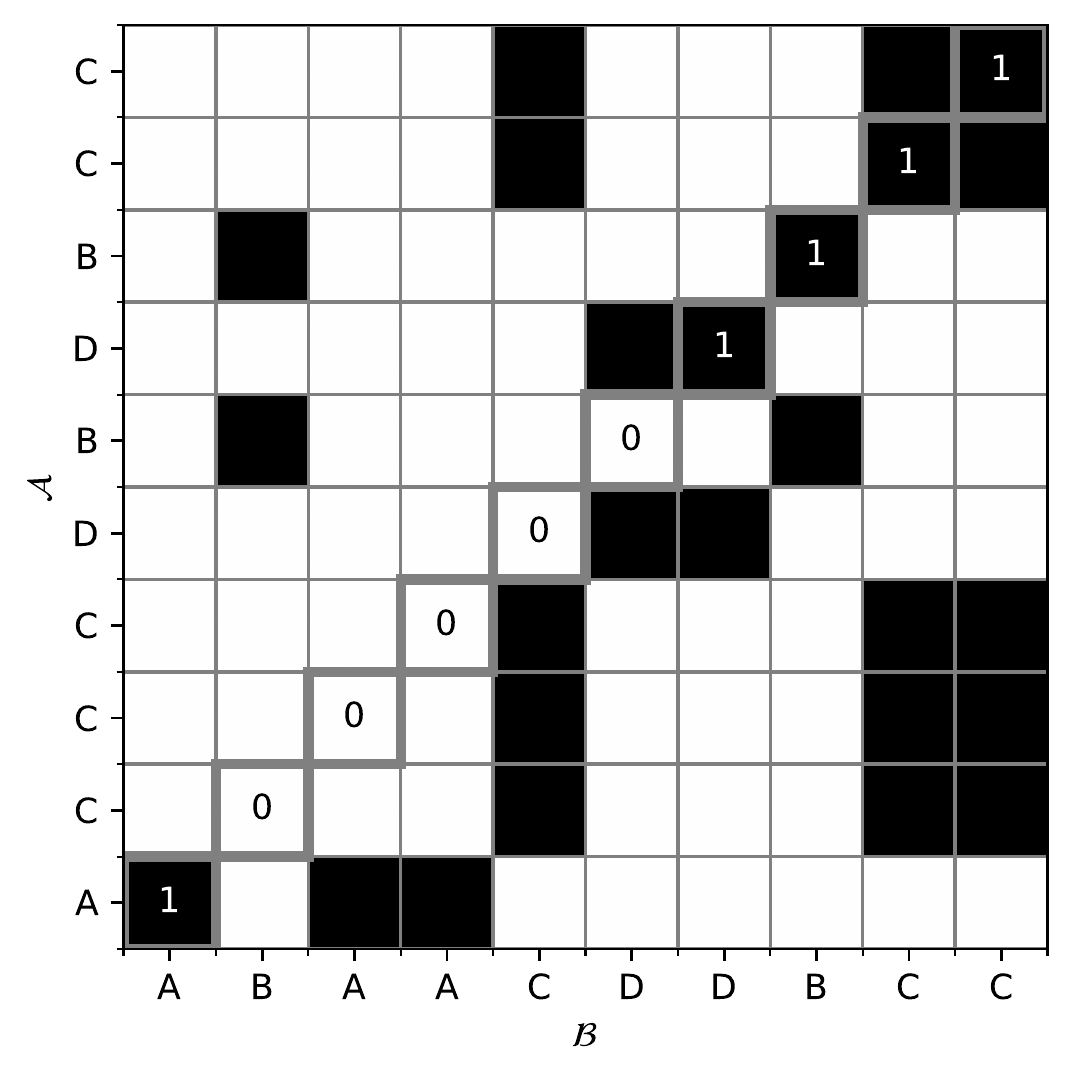}
\caption{The CRP of two sample time series with the diagonal line highlighted.}
\label{fig:CRP_sample}
\end{figure}

A simple example with two one-dimensional time series clarifies how we extract the input features and prediction targets. Consider the two time series $\A$ and $\B$ of 10 observations:
\begin{align*}
     \A &= \{A,B,A,A,C,D,D,B,C,C\}^\top, \\
     \B &= \{A,C,C,C,D,B,D,B,C,C\}^\top.
\end{align*}
Figure~\ref{fig:CRP_sample} depicts their CRP, i.e., $CRP_{(\A,\B)}$ (for simplicity computed with $k=\tau=1$, and $V=V'=10$). The diagonal line of the CRP is highlighted and includes the values of the recurrence states. The diagonal line shows that the behavior of $\A$ and $\B$ at timestamps between $1$ and $7$ to $10$ is synchronized, therefore at these timestamps the prediction targets are set to 1 (the actual value of the Heaviside function in \eqref{eq:CRij}).
By, for instance, setting $w = 3$, we aim at predicting $V-w = 7$ states of synchronization. 
The first prediction concerns the synchronization at epoch $w+1=4$, based on the $CRP_{(\N^3_3,\M^3_3)}$, that is, the CRP calculated from the first three observations of $\A$ and $\B$. The prediction of the synchronization at epoch $5$, is based on the CRP calculated on observations 2 to 4, i.e. on $CRP_{(\N^3_4,\M^3_4)}$. The procedure is repeated up to epoch $V-1=9$, where $CRP_{(\A^3_{9},\B^3_{9})}$, calculated from the observations 7 to 9, is used for predicting $\text{diag}(CRP_{(\A,\B)})_{10}$.

To practically implement the underlying DL model that maps each  $CRP_{(\N^w_{i} ,\M^w_{i})}$ input to its corresponding $\text{diag}\left(CRP_{(\N,\M)}\right)_{i+1}$ output, consider that each input consists of a matrix of zeros and ones that can be considered analogous to an image. Therefore we can rely on well-established classification models. In particular, we employ a Convolutional Neural Network (CNN). Such a neural network is well-suited for capturing the spatial relationships between the features in their input, which in our case correspond to the 0-1 features encoded in the entries of $CRP_{(\N^w_{i} ,\M^w_{i})}$. Note that in the CRP calculation $\N^w_{i}$ and $\M^w_{i}$ are $z$-score normalized before computing \eqref{eq:CRij}.
\begin{figure*}[ht]
\centering
\includegraphics[width=0.95\textwidth]{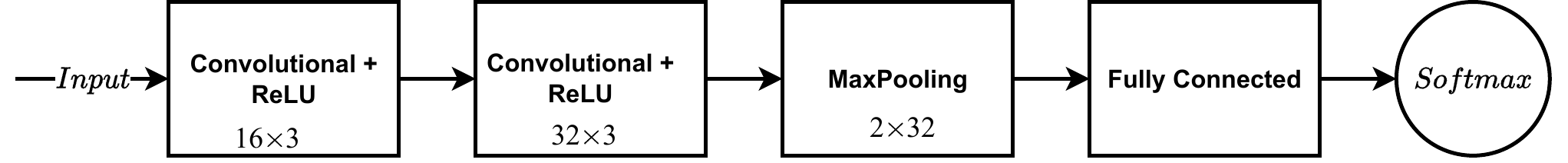}
\caption{Architecture of the proposed convolutional neural network.}
\label{fig:model}
\end{figure*}

We use a CNN architecture formed by two convolutional and one fully-connected layer, as illustrated in Figure~\ref{fig:model}. The neural network involves the typical blocks of the CNN architecture. The convolutional layers adaptively learn the spatial relationships of inputs, the Rectified Linear Unit (ReLU) activation introduces nonlinearity to the model, and the max-pooling layer provides down-sampling operations reducing the size of the feature map by extracting the maximum value in each patch from the input feature map. The current CNN is chosen based on a grid search over different network architectures, layers' types and sizes, aimed at maximizing the F1-score and showing the feasibility of our CRP-based DL approach. 

\section{Experiments}\label{S:Experiments}
\subsection{Data}\label{sec:Dataset}
Our analyses rely on daily adjusted closing prices and daily number of traded shares (volumes) for 12 representative constituents of the S\&P100 index in the period from December 31\textsuperscript{st}, 2014 to November 29\textsuperscript{th}, 2021 ($V= 1,741$ trading days). The data is retrieved from Yahoo Finance. These 12 stocks are selected based on their market capitalization and their market sector. For each sector we select the first two stocks of highest-but-comparable capitalization, a  practice well-supported by financial theory \cite{fama1993common}.
Market sectors provide a natural grouping for securities: analyses conducted at a sector level are a common practice for granting comparability and robustness of the results, as across market sectors the dynamics of economic variables are well-known to be asymmetric. Table~\ref{tb:data} lists our stock selection. 
Each stock is expressed as a trivariate time series consisting of daily prices, volumes, and returns. This way, CRPs express temporal similarities in joint terms of the price level, traded volume, and daily return, providing a generalized definition of similarity in time-series dynamics at a multivariate level.

For our bivariate analysis on two time series we have $(12^2-12)/2 = 66$ pairs of stocks.
For each stock pair, we use the first $70\%$ of the data for training ($V_{train} = 1,218$ days) and the last $30\%$ for testing ($V_{test} = 523$ days). As the future input instances should not affect the training process, the order of the input data during the training is fixed. The input and targets of the train data and the test data are, respectively,
\begin{align*}
\text{Inputs:}& \;\left\lbrace  CRP_{(\N^w_{i} ,\M^w_{i})} \right\rbrace_{i \in I}\text{,}\\
\text{Targets:}& \;\left\lbrace\text{diag}\left(CRP_{(\N,\M)}\right)_i \right\rbrace_{i \in T} \text{,}
\end{align*}
where $I = w,\dots,V_\text{train}$ and $T =w+1,\dots, V_\text{train}+1$ for the training set and $I = V_\text{train}+1,\dots,V-1$ and $T =V_\text{train}+2,\dots, V$ for the test set. 
We train the neural network once over the data for all the picks of the stock pairs. This pooled approach is a common practice in closely-related Machine Learning literature \cite[e.g.]{ntakaris2018benchmark,tran2018temporal} and supported e.g. by the empirical findings of \cite{Sirignano2019universal}, suggesting the existence of an universal price formation mechanism (model), and thus price dynamic, not specific for individual assets. In practice, the input and output data is the concatenation of the individual pairs' inputs-targets. For example, for a set window size $w$, for the train set the input-target data consists of $(V'_\text{train}-w) \times 66$ examples, that is $(V'_\text{train}-w) \times 66$ pairs of cross-recurrent matrices and (scalar) targets, where $V'_\text{train} = (V_\text{train}-\tau (k-1))$.
In the training phase, the training data is used to estimate the optimal weights of the CNN. The test data is then parsed to the estimated CNN and the quality of the network outputs is evaluated against the actual targets. Details are provided in the following two subsections.

\begin{table}
\caption{List of selected stocks.}
\centering
\begin{tabular}{lll}
\toprule
\multicolumn{1}{c}{Sector} & \multicolumn{1}{c}{Ticker} & \multicolumn{1}{c}{Stock name} \\ 
\midrule
Electronic Technology (ET) & INTC & Intel Corporation \\ 
Electronic Technology & QCOM & Qualcomm Inc. \\ 
Energy   Minerals (EM) & XOM & Exxon   Mobil Corporation \\ 
Energy   Minerals & CVX & Chevron   Corporation \\ 
Finance (F) & JPM & JP Morgan Chase \& Co. \\ 
Finance & V & Visa Inc. \\ 
Health   Technology (HT) & JNJ & Johnson   \& Johnson \\ 
Health   Technology & PFE & Pfizer,   Inc. \\ 
Retail Trade (RT) & HD & Home Depot, Inc. (The) \\ 
Retail Trade & WMT & Walmart Inc. \\ 
Technology   Services (TS) & MSFT & Microsoft   Corporation \\ 
Technology   Services & GOOG & Alphabet   Inc. \\ 
\bottomrule
\end{tabular}
\label{tb:data}
\end{table}

For the training of the CNN we adopt the ADAM optimizer with the following hyper-parameters: learning rate $0.01$ (reduced by a factor of $5$ every $40$ epochs), momentum parameters $0.9$ and $0.999$, batch size $128$ and epoch size $300$. Across the epochs we keep track of the F1-score on the validation set, which is set to the last 15\% portion of the training set. For our classification task we adopt the binary cross-entropy loss. As the target classes are unbalanced, the loss is weighted for the targets' class proportion. Details on the filter sizes, kernel sizes and the max pooling size are provided in Figure \ref{fig:model}.

With respect to the CRP computations, throughout our analyses the embedding dimension $k$ is set to 2 or 3 (estimated via FNN method) based on input type, and the delay parameter $\tau$ is set to $1$. Values $0.45$, $0.55$, $0.65$, and $0.75$ are used for the threshold $\varepsilon$. These hyper-parameters are selected according to the guidelines and discussion in \cite{schinkel2008selection} and \cite{wallot2019multidimensional}. The same values are applied for both the computation for the CRP related to the targets and the CRPs related to the inputs. 

In our experiments we consider two different choices for the window-size hyperparameter, namely $w=\{10,30, 50, 60, 80\}$ days.
With the above settings, $V=V'=1,741$ days, $V_\text{train} = V'_\text{train} =1,218$, and $V_\text{test} = V'_\text{test} = 523$ days. For $i = w,\dots,V-1$, $CRP_{(\N^w_i,\M^w_i)}$ are square matrices of size $w'=w$ and $CPR_{(\N,\M)}$ a square matrix of size $V$ on whose diagonal are found the relevant targets, i.e. $\text{diag}\left(CPR_{(\N,\M)}\right)_i$, $i = w+1,\dots,V$.

\subsection{Experiments Results}\label{subsec:experiments}

Stock pairs from the same sector or two different sectors with different co-movement behaviors can provide comprehensive experimental data to show the ability of the proposed method in predicting the state of synchronization.
To evaluate the performance of our proposed method, all pairs of stocks are used as the input of the method. We collect all pairs of stocks and for each pair, we follow the steps of the proposed method (Fig~\ref{fig:method}) to create the inputs and targets. We stack the input-target pair-specific data to create a single train and test set for all pairs.  

Tables \ref{tb:rez-p-V} and ~\ref{tb:rez-p-V-R}, show the performance of our proposed approach for all pairs of stocks using two types of input: (price, volume) and (price, return, volume) respectively. Given that the target classes are generally imbalanced, the preferred reference performance metric is the F1 score. Yet, we also include accuracy, precision, and recall to have a clearer overview of the classification performance. 
For robustness, we run our experiments over a range of values for the window-size $W$ and threshold $\varepsilon$ hyperparameters, a setup that further clarifies the effect of these hyperparameters on prediction performance.

Results for the (price, volume) time-series input are provided in Table~\ref{tb:rez-p-V}, results for the (price, return, volume) input in Table~\ref{tb:rez-p-V-R}.
In general, our results show that the task of predicting the state of synchronization is not only feasible, but, under our setup, quite satisfactory. Indeed our preferred performance F1 metric is as high as 84\%. Yet, as expected, the results appear to be sensible to the choice of the window size and threshold parameter. In particular, the performance metrics decrease in their values as the threshold parameter and the window size increase. This means that stricter $\varepsilon$-neighbourhoods are easier to predict and that the relevant information for the prediction of the synchronization state is found in the most recent instances of the CRP.
This suggests the existence of patterns in the data that are strongly indicative of close $\varepsilon$-neighbourhoods, for which the prediction is very satisfactory. I.e. the CNN detects clear patterns indicative of the fact that the day-ahead synchronization is likely to be very strong (the $\varepsilon$-neighbourhoods is tight), indeed, as $\varepsilon$ increases, the performance metric decrease, indicating that the model indeed detects strong evidence of \enquote{strong} day-ahead synchronization.
Regarding, the window size, Long-lagged CRP information appears to introduce noise in the system without providing any predictive gains, aligned with the intuition that further-in-time information is less and less related to the current state of the system and of little use for prediction.

Suspecting that the use of prices and returns might be redundant, since they are closely related to each other, we also run a second experiment involving volumes and returns only.
It is interesting to note that the inclusion of the returns does not seem to provide any advantage with respect to the (price, volume) input time series, but rather the opposite effect. 
It is indeed expected that the inclusion of further input variables complicates the patterns in the CRP chessboard so that under the same network architecture the performance metrics decrease.
Furthermore, and aligned with the above, in additional experiments here not reported, we included squared returns (as a gross measure of daily volatility) finding that they also appear to have a detrimental on the performance metrics and prediction task. This perhaps suggests that the network architecture needs to scale up with the complexity of the input data (number of time series) that reasonably induces more complex patterns in the CRP. 

\begin{table}
  \centering
  \caption{Performance measures on the test set  using (adjusted) price and volume as input variables.}
    \begin{tabular}{cccccc}
    $w$   & $\varepsilon$ & Accuracy & Precision & Recall & f1-score \\
    \midrule
    10    & 0.45  & 0.960  & 0.886 & 0.818 & \textbf{0.848} \\
    10    & 0.55  & 0.981 & 0.842 & 0.762 & 0.796 \\
    10    & 0.65  & 0.992 & 0.836 & 0.684 & 0.737 \\
    10    & 0.75  & 0.997 & 0.999 & 0.647 & 0.727 \\
    \midrule
    30    & 0.45  & 0.957 & 0.877 & 0.804 & 0.836 \\
    30    & 0.55  & 0.979 & 0.821 & 0.752 & 0.782 \\
    30    & 0.65  & 0.991 & 0.816 & 0.684 & 0.732 \\
    30    & 0.75  & 0.996 & 0.998 & 0.539 & 0.571 \\
    \midrule
    50    & 0.45  & 0.956 & 0.861 & 0.808 & 0.832 \\
    50    & 0.55  & 0.982 & 0.907 & 0.719 & 0.784 \\
    50    & 0.65  & 0.993 & 0.907 & 0.668 & 0.737 \\
    50    & 0.75  & 0.997 & 0.935 & 0.665 & 0.739 \\
    \midrule
    60    & 0.45  & 0.954 & 0.850  & 0.814 & 0.831 \\
    60    & 0.55  & 0.976 & 0.775 & 0.730  & 0.751 \\
    60    & 0.65  & 0.992 & 0.937 & 0.633 & 0.703 \\
    60    & 0.75  & 0.997 & 0.816 & 0.689 & 0.737 \\
    \midrule
    80    & 0.45  & 0.950  & 0.827 & 0.809 & 0.818 \\
    80    & 0.55  & 0.979 & 0.839 & 0.725 & 0.769 \\
    80    & 0.65  & 0.990  & 0.776 & 0.666 & 0.707 \\
    80    & 0.75  & 0.997 & 0.820  & 0.709 & 0.753 \\
    \bottomrule
    \end{tabular}%
    \label{tb:rez-p-V}
\end{table}%

\begin{table}
\centering
\caption{Performance measures on the test set  using (adjusted) price, volume and returns as input variables.}
    \begin{tabular}{cccccc}
    $w$   & $\varepsilon$ & Accuracy & Precision & Recall & f1-score \\
    \midrule
    10    & 0.45  & 0.946 & 0.858 & 0.802 & \textbf{0.827} \\
    10    & 0.55  & 0.973 & 0.810 & 0.733 & 0.765 \\
    10    & 0.65  & 0.988 & 0.776 & 0.694 & 0.728 \\
    10    & 0.75  & 0.995 & 0.795 & 0.664 & 0.711 \\
    \midrule
    30    & 0.45  & 0.943 & 0.845 & 0.800 & 0.820 \\
    30    & 0.55  & 0.971 & 0.789 & 0.742 & 0.763 \\
    30    & 0.65  & 0.987 & 0.760 & 0.678 & 0.711 \\
    30    & 0.75  & 0.996 & 0.998 & 0.601 & 0.667 \\
    \midrule
    50    & 0.45  & 0.938 & 0.826 & 0.785 & 0.804 \\
    50    & 0.55  & 0.971 & 0.795 & 0.731 & 0.758 \\
    50    & 0.65  & 0.987 & 0.758 & 0.667 & 0.702 \\
    50    & 0.75  & 0.995 & 0.998 & 0.503 & 0.505 \\
    \midrule
    60    & 0.45  & 0.938 & 0.823 & 0.788 & 0.804 \\
    60    & 0.55  & 0.967 & 0.752 & 0.735 & 0.743 \\
    60    & 0.65  & 0.987 & 0.756 & 0.654 & 0.692 \\
    60    & 0.75  & 0.996 & 0.955 & 0.595 & 0.657 \\
    \midrule
    80    & 0.45  & 0.933 & 0.805 & 0.788 & 0.796 \\
    80    & 0.55  & 0.967 & 0.755 & 0.730 & 0.742 \\
    80    & 0.65  & 0.989 & 0.893 & 0.616 & 0.677 \\
    80    & 0.75  & 0.995 & 0.998 & 0.548 & 0.586 \\
    \bottomrule
    \end{tabular}
    \label{tb:rez-p-V-R}
\end{table}

\section{Conclusion}\label{S:Conclusion}
Predicting the co-movement of two multidimensional time series is a relevant task for the financial industry that supports potential trading strategies based on their interrelationships.
This paper contributes to the literature by providing (i) a method relying upon the CRP to quantify the time series coupling over time, (ii) DL model for the prediction of the time series synchronization state, (iii) the use of a multidimensional time-series representation of the inputs involving prices, volumes and returns. Furthermore, (iv) we conduct extensive analyses on real stock market data from different sectors. The results show that the proposed setup can effectively predict the one-day-ahead synchronization of two-time series.
Interesting future research direction would be to investigate the applicability of such an approach to a high-frequency domain where the high-dimensional nature of the raw data may provide valuable information for analyzing and predicting the coupling in settings that are known to be characterized by high levels of noise. 

\section*{Acknowledgments}
The research received funding from the Independent Research Fund Denmark project DISPA (project No. 9041-00004), and the European Union’s Horizon 2020 research and innovation programme under the Marie Sk\l odowska-Curie project BNNmetrics (grant agreement No. 890690).

\bibliographystyle{apalike}
\bibliography{Bibliography}


\end{document}